1. INTRODUCTION

The longest integrations of planetary orbits are still well short of the age of the solar system. So far the full planetary system has been followed for 100 Myr (Sussman & Wisdom 1992), and the five outer planets for 1 Gyr (Wisdom & Holman 1991; hereafter WH). A semi-analytic secular perturbation theory has been used to follow the planetary system for 200 Myr (Laskar 1989, 1990) and shows very good agreement with direct integrations (Laskar et al. 1992, Sussman & Wisdom 1992). Progressively longer integrations have generally revealed interesting new phenomena, notably weak chaos in the orbits of the inner (Laskar 1989, 1990) and outer (Sussmann & Wisdom 1992) planets. This situation motivates the development of faster and more accurate integration methods.

Traditionally, long solar system integrations have used high-order multistep integration methods in Cartesian coordinates (see, for example, Quinn et al. 1991, hereafter QTD). However, substantial improvements in speed are possible using integration methods that are specifically designed for motion that is (a) Hamiltonian, and (b) nearly Keplerian; we call these mixed variable symplectic integrators and they are the subject of this paper.

The mixed variable symplectic (or MVS) integrators were introduced by WH, and also (independently and by different arguments) by Kinoshita et al. (1991; hereafter KYN). These derive their advantage by switching continually between Cartesian variables (wherein the perturbations are easy to evaluate) and Kepler elements (which make the solar part simple)—hence 'mixed variable'. The symplectic property (i.e., having certain Hamiltonian conservation laws built-in) helps control long-term errors. Saha & Tremaine (1992; hereafter Paper I) describe a startup technique ('warmup') that substantially improves the long-term accuracy of MVS integrators; a related technique ('symplectic correctors') is given by Wisdom et al. (1994; hereafter WHT). In this paper we generally follow KYN's methods of analysis but WH's algorithms.

A limitation of symplectic integrators so far is that they generally allow neither adaptive nor individual timesteps. Adaptive timesteps are indispensable for situations with close encounters or very high eccentricities, but for planetary and satellite orbits we do not miss them much. However, the requirement that all planets be followed with a common timestep is certainly undesirable, since planetary orbital periods range over three orders of magnitude. If the common timestep is dictated by the desired accuracy for Mercury, then for Pluto we may be paying for several unnecessary orders of magnitude of accuracy (see Fig. 2).

# Long-term planetary integration with individual time steps

Prasenjit Saha[1,2] and Scott Tremaine[1]

[1]Canadian Institute for Theoretical Astrophysics
McLennan Labs, University of Toronto,
60 St. George St., Toronto M5S 1A7, Canada
`tremaine@cita.utoronto.ca`

[2]Mount Stromlo and Siding Spring Observatories,
The Australian National University,
Canberra, ACT 0200, Australia
`saha@mso.anu.edu.au`

**Abstract:** We describe an algorithm for long-term planetary orbit integrations, including the dominant post-Newtonian effects, that employs individual timesteps for each planet. The algorithm is symplectic and exhibits short-term errors that are $O(\epsilon \Omega^2 \tau^2)$ where $\tau$ is the timestep, $\Omega$ is a typical orbital frequency, and $\epsilon \ll 1$ is a typical planetary mass in solar units. By a special starting procedure long-term errors over an integration interval $T$ can be reduced to $O(\epsilon^2 \Omega^3 \tau^2 T)$. A sample 0.8 Myr integration of the nine planets illustrates that Pluto can have a timestep more than 100 times Mercury's, without dominating the positional error. Our algorithm is applicable to other $N$-body systems.

The main contribution of this paper is to introduce an MVS integrator with individual timesteps. We describe our new algorithm in Sec. 3 after covering some operator formalism in Sec. 2. Section 4 has details of the equations of motion. In Sec. 5 we show how the leading order general relativistic corrections can be neatly incorporated. Section 6 has some numerical tests. Finally, in Sec. 7 we discuss some variations on our method.

The savings in computer time from adopting individual timesteps is modest in the case of the planetary system (a little over a factor of two), but still significant considering that solar system integrations often run for weeks or months of machine time. For other problems, however, (e.g. if the lunar orbit is integrated as well) the savings can be much larger.

## 2. LEAPFROG

Hamilton's equations can be written in terms of a Poisson bracket operator as

$$\frac{d}{dt} \equiv \{\ ,H\}. \tag{1}$$

which has the formal solution

$$e^{t\{\ ,H\}}. \tag{2}$$

An integration algorithm can be thought of as an approximate expression for the operator in (2).

A common variety of Hamiltonian is the sum of two (or more) parts, each of which is soluble in isolation. That is,

$$H = H_A + H_B, \tag{3}$$

where $e^{\{\ ,H_A\}}$ and $e^{\{\ ,H_B\}}$ are known. These two operators can be used as building blocks to construct approximations to (2). The obvious example is

$$H_A = \frac{p^2}{2m}, \quad H_B = V(\mathbf{r}), \tag{4}$$

as a kinetic energy and coordinate-dependent potential.

The solar system Hamiltonian can be expressed in the form (4), but it is more useful to take

$$H_A = H_{\text{Kep}}, \quad H_B = H_{\text{int}} \tag{5}$$

where $H_{\text{Kep}}$ is the kinetic energy plus solar potential and $H_{\text{int}}$ is the interaction energy between the planets (WH, KYN). Then $e^{t\{\ ,H_{\text{Kep}}\}}$ generates motion along unperturbed Kepler orbits, while $e^{t\{\ ,H_{\text{int}}\}}$ generates a change of momenta with the coordinates fixed.

A key ingredient in integration algorithms for such systems is the operator identity (see, for example, Yoshida 1993):

$$e^{\frac{1}{2}\tau\{\ ,H_A\}} e^{\tau\{\ ,H_B\}} e^{\frac{1}{2}\tau\{\ ,H_A\}} \equiv e^{\tau\{\ ,H_A+H_B+H_{\text{err}}\}} \tag{6}$$

where $H_{\text{err}}$ is a formal power series in $\tau$ starting at $O(\tau^2)$ and consisting of nested Poisson brackets of $H_A$ and $H_B$:

$$H_{\text{err}} = \frac{\tau^2}{12}\{\{H_A,H_B\},H_B + \tfrac{1}{2}H_A\} + O(\tau^4). \tag{7}$$

In general the series for $H_{\text{err}}$ does not converge[3], and is interpreted as an asymptotic series. The left side of (6) is recognized as one step of a generalized form of leapfrog, with $\tau$ being the timestep. The well-known result that leapfrog is second order follows from the fact that $H_{\text{err}}$ is $O(\tau^2)$. The expression (6) also reveals other useful properties:

(i) The integration errors are Hamiltonian; that is, the integration algorithm follows exactly the dynamics of a nearby 'surrogate' Hamiltonian

$$H + H_{\text{err}}. \tag{8}$$

This result must be interpreted cautiously since $H_{\text{err}}$ is only a formal series, but in practice, for small $\tau$, analyses based on the leading term in $H_{\text{err}}$ provide considerable insight.

(ii) For orbital motion, (i) suggests that the energy error is bounded and therefore the position errors grow as the integration time $T$, rather than as $T^2$ (as in most integration methods). Moreover, the rate of error growth can be estimated from $H_{\text{err}}$, as illustrated by KYN.

Now suppose that $H_B$ is $O(\epsilon)$ smaller than $H_A$, as in planetary motion if we identify $H_A$ and $H_B$ as in Eq. (5) (in this case $\epsilon$ is of order the planetary mass). Two further properties then follow.

---

[3] This is illustrated by the following example (J. Wisdom, private communication): consider the pendulum Hamiltonian, with $H_A = \frac{1}{2}p^2$ and $H_B = -\cos q$. The map (6) applied to $z = (q,p)$ is then simply the standard map (to within a shear). The standard map exhibits chaos, which is incompatible with motion in an autonomous one degree-of-freedom Hamiltonian $H_A + H_B + H_{\text{err}}$.

(iii) Since $H_{\rm err}$ consists entirely of nested Poisson brackets of $H_{\rm A}$ and $H_{\rm B}$, $H_{\rm err}$ will be of $O(\epsilon\tau^2)$.

(iv) The long term errors can be reduced from $O(\epsilon\tau^2 T)$ to $O(\epsilon^2\tau^2 T)$ by special starting procedures (Paper I—see also WHT). The reason is that $H_{\rm err}$ has no secular terms at $O(\epsilon)$,[4] so that to this order the relation between actions and frequencies is the same in the actual and surrogate systems. Thus any $O(\epsilon\tau^n T)$ error terms come from a difference in the values for the actions in the actual and surrogate systems—this difference of course giving rise to a constant frequency error and hence a linearly growing position error. If the difference between the surrogate and exact action values can be removed (to leading order in $\epsilon$) by a suitable small alteration in the initial conditions, all $O(\epsilon\tau^n T)$ error can be suppressed.

By concatenating leapfrog steps one can produce higher order integrators; for example, three consecutive leapfrog steps with timesteps in the ratio $1 : -2^{\frac{1}{3}} : 1$ amount to a single step of a fourth order integrator. Yoshida (1990) shows that arbitrarily high orders are possible. With suitable modifications, properties (i)–(iv) carry over to higher order integrators.

In this paper, we develop an integration algorithm with individual timesteps for each planet. The idea is to apply Eq. (6) recursively, replacing the operator $e^{t\{\ ,H_{\rm B}\}}$ by a more complicated operator that itself involves a leapfrog step. The arguments leading to properties (i)–(iv) are not affected by this change.

### 3. LEAPFROG WITH INDIVIDUAL TIMESTEPS

Suppose that the Hamiltonian for a system of planets is split into $H_{\rm Kep}$ and $H_{\rm int}$ as in Eq. (5). The details of what is inside $H_{\rm Kep}$ and $H_{\rm int}$ we leave for Sec. 4. Now imagine two clocks $K$ and $I$, associated with $H_{\rm Kep}$ and $H_{\rm int}$ respectively. These resemble the clocks used in chess tournaments in that only one of them is running at a given time; when the 'Kepler clock' $K$ is running, each of the planets moves along its osculating Kepler orbit, and the interplanetary interactions are turned off; when the 'interaction clock' $I$ is running all the coordinates stay fixed while the momenta change according to $H_{\rm int}$. Thus an interval $\tau$ of $K$ or $I$ corresponds to the operators $e^{\tau\{\ ,H_{\rm Kep}\}}$ or $e^{\tau\{\ ,H_{\rm int}\}}$ respectively. A single leapfrog step of length $\tau$ can be written in pseudocode as the following procedure:

$$\begin{array}{l}\langle\text{Advance } K \text{ by } \tfrac{1}{2}\tau\rangle \\ \langle\text{Advance } I \text{ by } \tau\rangle \\ \langle\text{Advance } K \text{ by } \tfrac{1}{2}\tau\rangle\end{array} \quad (9)$$

---

[4] The exposition in Sec. 3 of Paper I is flawed in that it shows only that the error Hamiltonian has no secular terms at $O(\epsilon\tau^2)$ (P.-V. Koseleff, private communication); however, the argument in that Section readily extends to show that there are no secular terms of order $O(\epsilon\tau^n)$ for any $n$.

We can write a sequence of leapfrog steps, each of size $\tau$, in terms of the $K$ and $I$ clocks as follows.

$$\begin{array}{l}\langle\text{Advance } K \text{ by } \tfrac{1}{2}\tau\rangle \\ \textbf{loop} \\ \quad \langle\text{Advance } I \text{ by } \tau\rangle \\ \quad \langle\text{If output is desired then } \textbf{exit loop}\rangle \\ \quad \langle\text{Advance } K \text{ by } \tau\rangle \\ \textbf{end loop} \\ \langle\text{Advance } K \text{ by } \tfrac{1}{2}\tau\rangle\end{array} \quad (10)$$

Note that $K$ and $I$ show the same time at the start and end, but in between they are not synchronized.

Now we propose a generalization to individual timesteps. First we assign each planet its own timestep (which normally does not vary during the integration). Assume that the planets are indexed from innermost outwards as $1..N$, planet $i$ has timestep $\tau_i$, and we restrict $\tau_{i+1}$ to be an integer multiple of $\tau_i$. We assume that $H_{\rm Kep}$ and $H_{\rm int}$ can be written in the form

$$H_{\rm Kep} = \sum_{i=1}^{N} H_{{\rm Kep},i}, \qquad H_{\rm int} = \sum_{i=1}^{N} H_{{\rm int},i}. \quad (11)$$

Here $H_{{\rm Kep},i}$ is the Kepler Hamiltonian for the two-body system consisting of the sun and planet $i$, and $H_{{\rm int},i}$ is the potential energy arising from the interaction of planet $i$ with planets $i+1$ to $N$. These Hamiltonians have the properties

$$\{H_{{\rm Kep},i}, H_{{\rm Kep},j}\} = 0, \quad \{H_{{\rm int},i}, H_{{\rm int},j}\} = 0, \quad (12a)$$

and

$$\{H_{{\rm Kep},i}, H_{{\rm int},j}\} = 0, \quad \text{for } j > i. \quad (12b)$$

(In Sec. 7 we discuss algorithms for the case where Eq. 12b does not apply.)

We can now assign clocks $K_i$ and $I_i$ to each planet: advancing a Kepler clock $K_i$ by $\tau$ corresponds to the operator $e^{\tau\{\cdot,H_{\text{Kep},i}\}}$, that is, to advancing planet $i$ along its osculating Kepler orbit; advancing an interaction clock $I_i$ corresponds to the operator $e^{\tau\{\cdot,H_{\text{int},i}\}}$, that is, to changing momenta according to the interactions between planet $i$ and all planets $j > i$ (note that the clock $I_N$ does nothing). Equations (12a,b) imply that the order of advancing any two Kepler clocks $K_i$, $K_j$ can be reversed without affecting the result (i.e. the corresponding operators commute); the same is true for any two interaction clocks $I_i$, $I_j$, and also for $K_i$, $I_j$ when $i < j$.

The general idea of our algorithm is to interleave advances of the various clocks such that (i) at the end of the integration each of the clocks has been advanced by the same amount; (ii) the integration is reversible, that is, if the algorithm is applied to the time-reversed final state we recover the time-reversed initial state; (iii) the clocks remain as close to synchronization as possible. A suitable algorithm is expressed by the following recursive procedure:

$$
\begin{aligned}
&\textbf{procedure } \text{TICK}(i) \\
&\quad \langle \text{Advance } K_i \text{ by } \tfrac{1}{2}\tau_i \rangle \\
&\quad \langle \text{Advance } I_i \text{ by } \tau_i \rangle \\
&\quad \textbf{if } i > 1 \\
&\quad\quad \textbf{loop } \tau_i/\tau_{i-1} \textbf{ times} \\
&\quad\quad\quad \textbf{call } \text{TICK}(i-1) \\
&\quad\quad \textbf{end loop} \\
&\quad \textbf{end if} \\
&\quad \langle \text{Advance } K_i \text{ by } \tfrac{1}{2}\tau_i \rangle \\
&\textbf{end TICK}
\end{aligned}
\qquad (13)
$$

The order of the step that advances $I_i$ and the loop that calls TICK $(i-1)$ can be reversed without affecting the result; bearing this in mind it is easy to see that the algorithm is time-reversible. To advance the integration of the $N$-body system by $\tau_N$ one simply calls TICK $(N)$.

However (13) is not the most efficient form of the algorithm, because it often splits an advance of a $K_i$ clock by $\tau_i$ into two successive advances by $\tfrac{1}{2}\tau_i$. The form we actually implement is an equivalent non-recursive

version:

$$
\begin{aligned}
&\langle \text{Advance all the } K_i \text{ by } \tfrac{1}{2}\tau_i \rangle \\
&\textbf{loop} \\
&\quad \langle \text{For any } i \text{ where } K_i \text{ has changed more recently than } I_i, \\
&\quad\quad \text{advance } I_i \text{ by } \tau_i \rangle \\
&\quad \langle \text{If the } I_i \text{ are all equal and output is desired then } \textbf{exit} \\
&\quad\quad \textbf{loop} \rangle \\
&\quad \textbf{loop for } i = 1..N \\
&\quad\quad \langle \text{If } i = 1, \text{ or } K_i + \tfrac{1}{2}\tau_i \le K_{i-1}, \text{ advance } K_i \text{ by } \tau_i \rangle \\
&\quad \textbf{end loop} \\
&\textbf{end loop} \\
&\langle \text{Advance all the } K_i \text{ by } \tfrac{1}{2}\tau_i \rangle
\end{aligned}
\qquad (14)
$$

Note that although the clock $I_N$ does nothing, it must be monitored along with the other interaction clocks. It may be helpful to follow the steps in the algorithm by hand in a few simple cases to see how it works.

As an example, consider the case of two planets, with $\tau_2 = 2\tau_1 \equiv 2\tau$. Since $I_2$ does nothing we may write $I_1 \equiv I$. A single step (of duration $2\tau$) is:

$$
\begin{aligned}
&\langle \text{Advance } K_1 \text{ by } \tfrac{1}{2}\tau \text{ and } K_2 \text{ by } \tau \rangle \\
&\langle \text{Advance } I \text{ by } \tau \rangle \\
&\langle \text{Advance } K_1 \text{ by } \tau \rangle \\
&\langle \text{Advance } I \text{ by } \tau \rangle \\
&\langle \text{Advance } K_1 \text{ by } \tfrac{1}{2}\tau \text{ and } K_2 \text{ by } \tau \rangle
\end{aligned}
\qquad (15)
$$

It is straightforward to derive the error Hamiltonian for this case:

$$
H_{\text{err}} = \frac{\tau^2}{12}\left[(1I,I) + \tfrac{1}{2}(1I,1) + 4(2I,I) + 2(2I,2) + 4(2I,1)\right] + O(\tau^4), \qquad (16)
$$

where the symbols $(ij,k)$ represent nested Poisson brackets; for example, $(2I,1) \equiv \{\{H_{\text{Kep},2}, H_{\text{int}}\}, H_{\text{Kep},1}\}$. The first two terms in Eq. (16) represent the error arising from the leapfrog step of length $\tau$ in planet 1 (cf. Eq. 7); the second two arise from the leapfrog step of length $2\tau$ in planet 2; and the final term is associated with the presence of both planets.

Notice that when the interaction between two planets $i$ and $j$ is computed, the clocks $K_i$ and $K_j$ are generally at different epochs. Second-order accuracy is still achieved because on average $K_i$ is behind $K_j$ as much as it is ahead of $K_j$. Accuracy could be improved by synchronizing the $K_i$ clocks when the interactions are to be computed. One way to do this would be to run all the $K_{j\neq i}$ clocks until they synchronize with $K_i$, advance $I_i$ by $\tau_i$, then run all the $K_{j\neq i}$ clocks back to their previous settings. This strategy is not useful for solar system integrations, because moving the planets back and forth along Kepler orbits takes too much computer time to be worth the gain in accuracy (it would be useful in a system where advancing $K_i$ took much less computer time than advancing $I_i$). A better plan is to replace the Kepler orbit, for the purpose of the synchronization only, by an approximation (we might call this 'symplectic interpolation'). Even a crude approximation, provided it is symplectic, improves the accuracy substantially. We have found that a simple rotation in the invariable plane by a preset amount works well. That is, we replace advances of any $I_i$ clock with the sequence of steps:

$$\begin{aligned}&\langle\text{Evaluate } \phi_j = \bar{n}_j(K_i - K_j) \text{ for } j > i,\ \bar{n}_j \text{ being the mean}\\ &\quad\text{motion of planet } j \text{ at the start of the integration}\rangle\\ &\langle\text{Rotate each planet } j > i \text{ in the invariable plane by } \phi_j\rangle\\ &\langle\text{Advance } I_i \text{ by } \tau_i\rangle\\ &\langle\text{Rotate each planet } j > i \text{ in the invariable plane by } -\phi_j\rangle\end{aligned} \qquad (17)$$

As described so far the algorithm leaves errors of $O(\epsilon\tau_1^2 T)$. To reduce long-term errors further to $O(\epsilon^2\tau_1^2 T)$ we use a special starting procedure. The general idea is to change adiabatically (i.e., over a time much longer than the orbital time but much shorter than the total integration time) from a much more accurate integration procedure to the one we will actually use. This procedure ensures that the error Hamiltonian $H_{\text{err}}$ grows slowly so that the actions are unchanged (cf. point (iv) in Sec. 2), which removes the leading source of long-term error. In Paper I we recommended starting with a very small timestep and then gradually increasing the timestep to its final value. (With individual timesteps, this would require starting with all the $\tau_i$ scaled down and then gradually scaling them up again, always maintaining a fixed ratio between the $\tau_i$.) This procedure did not produce the hoped-for gain in accuracy with our new algorithm—apparently because changing the timestep causes the system to sweep through artificial resonances (see Wisdom & Holman 1992), which are prominent because of the large timesteps used by some of the planets. A slightly more subtle startup procedure works better:

$$\begin{aligned}&\langle\text{Integrate backward (or forward) for a large number of orbits}\\ &\quad\text{with the } \tau_i \text{ all scaled to small values (by at least } \sim\sqrt{\epsilon}\text{), while}\\ &\quad\text{gradually reducing the interaction strengths to zero}\rangle\\ &\langle\text{Integrate forward (or backward) to the starting point with}\\ &\quad\text{the regular timesteps } \tau_i, \text{ while gradually reviving the inter-}\\ &\quad\text{action strengths to the correct values}\rangle\end{aligned} \qquad (18)$$

During the first part $H_{\text{err}}$ is negligible (because the timesteps are small); during the second part $H_{\text{err}}$ is initially negligible (because the interplanetary interactions are small and the integrator follows Kepler orbits exactly) and then grows slowly, as required for adiabatic invariance of the actions. Hereafter, we refer to this procedure as a 'warm start' or 'warmup'. Its computational overhead is small compared to the main integration.

The warmup technique is based on eliminating any $O(\epsilon)$ contribution from $H_{\text{err}}$ to the actions. The symplectic corrector approach of WHT is to annihilate the $O(\epsilon)$ part of $H_{\text{err}}$ altogether with a canonical transformation. WHT apply the transformation to the initial conditions, integrate (using equal timesteps) as usual, and then apply the inverse transformation whenever output is desired. The difference between symplectic correctors and warmup is analogous to the difference between perturbation theory and averaging. As one might expect, both methods are equally good at controlling long-term errors; symplectic correctors have the advantages that they also remove short-term oscillatory errors at $O(\epsilon)$, and that they can be extended to higher order in $\epsilon$. They are, however, more complicated to implement, especially if extended to higher orders or to individual timesteps.

### 4. EQUATIONS OF MOTION

The material in this section mostly follows WH's discussion; however, we include a little more detail on the calculation of time-evolution under $H_{\text{int}}$, and skip over some other points not essential in our context.

The MVS integrators expect a Hamiltonian expressed as a sum of Kepler terms of the type $p^2/2m - \mu/r$ and interaction terms of the type $V(\mathbf{r})$; moreover, for the integrator to work efficiently the interaction terms associated with each body must be much smaller than its Kepler term. These requirements necessitate a special set of variables (well known in classical perturbation theory and for the same reasons), the Jacobi variables. Heliocentric variables will not do because then the Hamiltonian does not have the right form, and barycentric variables will not do because the sun's motion is not dominated by a Kepler part. To convert to Jacobi coordinates one orders the planets (inner-to-outer being usual and probably best, but not essential) and reckons the coordinates of each planet from the barycenter of the sun and all the previous ones.

We use Gaussian units: the solar mass is unity and $k^2$ is the gravitational constant. The planets have masses $m_i$ and heliocentric coordinates $\mathbf{r}_i$ (in this section the dummy indices $i$ and $j$ always range from 1 to $N$). We first define cumulative masses $\sigma_i$, and renormalized masses and gravitational constants $\tilde{m}_i$ and $\mu_i$:

$$\begin{aligned}\sigma_i &= \sigma_{i-1} + m_i, \quad \sigma_0 = 1, \\ \tilde{m}_i &= \frac{\sigma_{i-1}}{\sigma_i} m_i, \\ \mu_i &= \frac{\sigma_i}{\sigma_{i-1}} k^2.\end{aligned} \quad (19)$$

The Jacobi coordinates $\tilde{\mathbf{r}}_i$ are then defined by

$$\tilde{\mathbf{r}}_i = \mathbf{r}_i - \frac{1}{\sigma_{i-1}} \sum_{j<i} m_j \mathbf{r}_j, \quad (20a)$$

which has the inverse

$$\mathbf{r}_i = \tilde{\mathbf{r}}_i + \sum_{j<i} \frac{m_j \tilde{\mathbf{r}}_j}{\sigma_j}, \quad (20b)$$

If we add to the set $\tilde{\mathbf{r}}_1..\tilde{\mathbf{r}}_N$ the position ($\tilde{\mathbf{r}}_0$, say) of the barycenter of the whole system (in some inertial frame), then the momenta ($\tilde{\mathbf{p}}_0..\tilde{\mathbf{p}}_N$, say) conjugate to $\tilde{\mathbf{r}}_0..\tilde{\mathbf{r}}_N$ have the simple interpretation that $\tilde{\mathbf{p}}_0$ is the total momentum and

$$\tilde{\mathbf{p}}_i = \tilde{m}_i \tilde{\mathbf{v}}_i, \quad \text{where} \quad \tilde{\mathbf{v}}_i = \frac{d\tilde{\mathbf{r}}_i}{dt}. \quad (21)$$

The canonical set $\tilde{\mathbf{r}}_1..\tilde{\mathbf{r}}_N, \tilde{\mathbf{p}}_1..\tilde{\mathbf{p}}_N$ is collectively known as the Jacobi variables; $\tilde{\mathbf{r}}_0$ is ignorable and we disregard $\tilde{\mathbf{r}}_0, \tilde{\mathbf{p}}_0$ from now on.

The advantage of transforming to Jacobi variables is that in the barycentric frame

$$\text{Kinetic energy} = \sum_i \frac{\tilde{p}_i^2}{2\tilde{m}_i}. \quad (22)$$

For a derivation see Plummer (1960). WH interpret the transformation in an interesting way as a matrix diagonalization. In view of (21), we can write the full Hamiltonian as

$$H = H_{\text{Kep}} + H_{\text{int}} + H_{\text{misc}} \quad (23a)$$

where (cf. Eq. 11)

$$\begin{aligned}H_{\text{Kep}} &= \sum_i H_{\text{Kep},i}, \quad \text{where} \quad H_{\text{Kep},i} = \frac{\tilde{p}_i^2}{2\tilde{m}_i} - \mu_i \frac{\tilde{m}_i}{\tilde{r}_i}, \\ H_{\text{int}} &= \sum_i H_{\text{int},i}, \quad \text{where} \quad H_{\text{int},i} = H_{\text{dir},i} + H_{\text{indir}} \delta_{i1}, \\ H_{\text{dir},i} &= -k^2 \sum_{j>i} \frac{m_i m_j}{|\mathbf{r}_i - \mathbf{r}_j|}, \\ H_{\text{indir}} &= k^2 \sum_i m_i \left( \frac{1}{\tilde{r}_i} - \frac{1}{r_i} \right).\end{aligned} \quad (23b)$$

The designations $H_{\text{dir}}$ and $H_{\text{indir}}$ refer to the direct and indirect parts in the usage of WH; this is similar but not identical to the traditional usage. $H_{\text{misc}}$ denotes any other physical effects. The most interesting of these is the general relativistic correction, which we will approximate with a post-Newtonian Hamiltonian $H_{\text{PN}}$ to be discussed in the next section. Besides this we include a term $H_{\text{lun}}$ that represents the attraction of the sun on the quadrupole moment of the Earth-Moon system (see QTD for details). Thus we have

$$H_{\text{misc}} = H_{\text{PN}} + H_{\text{lun}}. \quad (23c)$$

Other effects such as solar oblateness, asteroids, and the galactic tidal field are thought to be $\lesssim 10^{-10}$ of the Kepler part (see QTD), and we neglect them.

Now we consider time evolution under $H_{\text{Kep}}$ and $H_{\text{int}}$.

$H_{\text{Kep}}$ of course generates evolution along a Kepler orbit. Computing this involves (implicitly or explicitly) transforming from Cartesian positions and velocities to orbital elements, incrementing the mean anomaly by the appropriate amount, and then changing back to Cartesian variables. As advocated by WH, this process is efficiently encapsulated in Gauss's $f$ and $g$ functions.

Now consider $H_{\rm dir}$. It is convenient first to compute the impulse in heliocentric variables and then to transform to Jacobi variables. The accelerations due to $H_{{\rm dir},i}$ are

$$\begin{aligned}
\frac{d\mathbf{v}_i}{dt}\bigg|_{{\rm dir},i} &= -k^2 \sum_{j\neq i} m_j \frac{(\mathbf{r}_i - \mathbf{r}_j)}{|\mathbf{r}_i - \mathbf{r}_j|^3}, \\
\frac{d\tilde{\mathbf{v}}_i}{dt}\bigg|_{{\rm dir},i} &= \frac{d\mathbf{v}_i}{dt}\bigg|_{{\rm dir},i} - \frac{1}{\sigma_{i-1}} \sum_{j<i} m_j \frac{d\mathbf{v}_j}{dt}\bigg|_{{\rm dir},i}.
\end{aligned} \quad (24)$$

Evolution under $H_{\rm indir}$ is more complicated. Because $H_{\rm indir}$ mixes $r_i$ and $\tilde{r}_i$, it cannot be usefully be split into a sum of contributions associated with each planet. The acceleration due to $H_{\rm indir}$ is

$$\frac{d\tilde{\mathbf{v}}_i}{dt}\bigg|_{\rm indir} = -\mu_i \left( \frac{\tilde{\mathbf{r}}_i}{\tilde{r}_i^3} - \frac{\mathbf{r}_i}{r_i^3} - \frac{1}{\sigma_i} \sum_{j>i} m_j \frac{\mathbf{r}_j}{r_j^3} \right). \quad (25)$$

We lump $H_{\rm indir}$ in with $H_{{\rm int},1}$ for the algorithm, and thus compute Eq. (25) whenever the $I_1$ clock is advanced. This procedure does not produce a bottleneck, because all the $d\tilde{\mathbf{v}}_i/dt|_{\rm indir}$ can be computed in $O(N)$ operations, while the $d\mathbf{v}_i/dt|_{{\rm dir},i}$ computations require $O(N^2)$ operations. We should also mention the standard practice of rewriting differences such as $\tilde{\mathbf{r}}_i/\tilde{r}_i^3 - \mathbf{r}_i/r_i^3$ in (25) to be less sensitive to roundoff error—see, for example, the discussion of Encke's method in Danby (1988).

## 5. POST-NEWTONIAN CORRECTIONS

General relativistic effects in planetary motion have fractional amplitude of order $k^2/c^2r \sim 10^{-8}$ at $r=1$ AU. Neglecting corrections that are smaller still by $O(k^2/c^2r)$ or by $O(m)$, the relativistic effects can be expressed through the post-Newtonian Hamiltonian (see Landau & Lifshitz 1975)[5]:

$$H_{\rm PN} = \frac{1}{c^2} \sum_i \left( \frac{\mu_i^2 \tilde{m}_i}{2\tilde{r}_i^2} - \frac{\tilde{p}_i^4}{8\tilde{m}_i^3} - \frac{3\mu_i \tilde{p}_i^2}{2\tilde{m}_i \tilde{r}_i} \right); \quad (26)$$

---

[5] This form for $H_{\rm PN}$ assumes that the metric has the form $ds^2 = [1 - 2k^2/(c^2r)]dt^2 - [1 + 2k^2/(c^2r)]d\mathbf{x}^2 + O(c^{-4})$, consistent with the isotropic or harmonic form of the Schwarzschild metric. This is the metric recommended by the IAU, but older papers (e.g. Brouwer & Clemence 1961) often use the "standard" form of the Schwarzschild metric, which yields different equations of motion.

note that the distinction between barycentric, heliocentric, and Jacobi coordinates, and between $\mu_i$ and $k^2$ or $\tilde{m}_i$ and $m_i$, is negligible to the accuracy we are considering.

$H_{\rm PN}$ mixes coordinate and momentum dependencies and hence is not easily decomposed in the form (3). The usual practice has been to replace it by an *ad hoc* potential $U_{\rm PN}$ designed to mimic the most important effects (Nobili & Roxburgh 1986, Paper I, Sussman & Wisdom 1992). Alternatively one could replace $H_{\rm PN}$ by its average along Kepler orbits $\bar{H}_{\rm PN}$. Neither of these approaches is entirely satisfactory. The reason is that the difference between $H_{\rm PN}$ and $U_{\rm PN}$ or $\bar{H}_{\rm PN}$ usually contains a secular part; this does not noticeably affect the perihelion precession, but gives the orbital frequency an error of order the post-Newtonian effect itself (this could perhaps be alleviated by some special starting procedure analogous to warmup, though it is not obvious how).

However there is a simple extension of the MVS method that easily accommodates $H_{\rm PN}$, and to which we devote the rest of this section. We rearrange Eq. (26) as

$$H_{\rm PN} = \sum_i \left( \alpha_i H_{{\rm Kep},i}^2 + \beta_i/\tilde{r}_i^2 + \gamma_i \tilde{p}_i^4 \right) \quad (27{\rm a})$$

where

$$\alpha_i = 3/(2\tilde{m}_i c^2), \qquad \beta_i = -\mu_i^2/c^2, \qquad \gamma_i = -1/(2\tilde{m}_i^3 c^2). \quad (27{\rm b})$$

This is similar to the form (3)—except that there are now three terms, each of which is integrable in isolation—and we may thus compute time evolution as follows.

Each $\alpha_i H_{{\rm Kep},i}^2$ we combine with the corresponding $H_{{\rm Kep},i}$ in Eq. (23b). Then advancing the $K_i$ clock by $\tau_i$ is redefined as the operator

$$\exp\left(\tau_i\{\ , H_{{\rm Kep},i} + \alpha_i H_{{\rm Kep},i}^2\}\right). \quad (28)$$

Under the operator (28) $H_{{\rm Kep},i}$ is conserved and equals $-\tfrac{1}{2}\mu_i \tilde{m}_i/a_i$ ($a_i$ being the osculating semi-major axis in Jacobi variables), so (28) is equivalent to

$$\exp\left(\tau_i'\{\ , H_{{\rm Kep},i}\}\right), \qquad \tau_i' = \left(1 - \frac{3\mu_i}{2c^2 a_i}\right)\tau_i. \quad (29)$$

Thus, to incorporate the $\alpha_i H_{{\rm Kep},i}^2$ terms, we merely rescale the time argument passed to the $f$ and $g$ functions. The $\beta_i/\tilde{r}_i^2$ are trivial to deal with—we absorb them inside $H_{\text{indir}}$. The $\gamma_i p_i^4$ terms we incorporate through leapfrog operators that evolve under a $\tilde{p}_i^4$ term before and after advancing a $K_i$ clock.

Because $H_{\text{PN}}$ changes the momentum dependence of the total Hamiltonian, we no longer have $\tilde{\mathbf{p}}_i = \tilde{m}_i d\tilde{\mathbf{r}}_i/dt$ (Eq. 21), so the expressions for accelerations in Sec. 4 have to be modified. A simple way to incorporate the necessary modifications is as follows. We redefine $\tilde{\mathbf{v}}_i$ to be a pseudo velocity:

$$\tilde{\mathbf{v}}_i \neq \frac{d\tilde{\mathbf{x}}_i}{dt}, \quad \tilde{\mathbf{v}}_i = \frac{\tilde{\mathbf{p}}_i}{\tilde{m}_i}. \tag{30}$$

Writing out half of Hamilton's equations for the full Hamiltonian of Eq. (23) we have

$$\frac{d\tilde{\mathbf{x}}_i}{dt} = \tilde{\mathbf{v}}_i \left[ 1 - \frac{1}{c^2}\left( \frac{\tilde{v}_i^2}{2} + \frac{3\mu_i}{\tilde{r}_i} \right) \right], \tag{31}$$

relating the true velocity and pseudo velocity. The expressions for accelerations in Sec. 4 become valid again if the $\tilde{\mathbf{v}}_i$ are interpreted as pseudo velocities. Operationally, this means that we have to transform the initial velocities to pseudo velocities by solving Eq. (31) for the $\tilde{\mathbf{v}}_i$, carry out the integration in terms of the pseudo velocities and then transform back to true velocities to output results.

In our implementation, the relativistic corrections consume less than 5% of the computing time.

## 6. SAMPLE INTEGRATIONS

We have implemented the individual time-step algorithm for nine planets. After some experimentation, we picked timestep ratios of $1 : 2 : 2 : 4 : 8 : 8 : 64 : 64 : 256$ for Mercury, Venus,...,Pluto, which makes the longitudes roughly equally accurate for the planets. Accuracy of $\sim 1$ arcsec per century (or about 1 radian in 20 Myr) requires the smallest timestep to be about one week, assuming a warmup is used at the start. Symplectic interpolation was used as described in Sec. 4 to synchronize the Kepler clocks before advancing the interaction clocks. The computer time required is quite implementation and compiler dependent, but our code takes about 15 sec for each kyr of integration on a 50 MHz Sparc-10; if all the timesteps are reduced to equal Mercury's our code takes about 35 sec per kyr. The latter case corresponds roughly to the Sussman & Wisdom (1992) integration after refinements by WHT—the main difference is that the other authors use a more approximate form for the post-Newtonian part. At the 1 arcsec per century level the new algorithm is about an order of magnitude faster than the 12th order symmetric multistep integrator in Quinlan & Tremaine (1990), but of course for high-enough accuracy requirements the latter would be more efficient (in the absence of roundoff errors).

Here we check our implementation, with and without individual timesteps, against the 12th order symmetric multistep integrator. For the symplectic integrator we set Mercury's timestep to be $7\frac{1}{32}$ days; with the other timesteps in the ratios above a single cycle then spans 1800 days. For the warmup we first integrated back 5000 yr with all the timesteps reduced 32-fold, while gradually reducing the interaction strengths to zero, then integrated forward to the starting point while gradually reviving the interactions again. For the multistep integrator we used a $\frac{1}{2}$ day timestep, which should be much more accurate than the symplectic integrator. In Fig. 1 the symplectic integration has individual time-steps; in Fig. 2 the symplectic timesteps are all reduced to equal Mercury's.

It is straightforward to estimate the time saved by integrating with individual timesteps. Suppose that advancing the Kepler clock $K_i$ takes computer time $\Delta t_K$, and that advancing the interaction clock $I_i$ takes time $(N-i)\Delta t_I$ (the number of interactions to be calculated equals the number of planets outside planet $i$). Then the ratio $r$ of computer time to integration time is

$$r = \sum_{i=1}^{N} \left[ \frac{\Delta t_K}{\tau_i} + (N-i)\frac{\Delta t_I}{\tau_i} \right]. \tag{32}$$

If a common timestep is used this expression simplifies to

$$r_{\text{c}} = N\frac{\Delta t_K}{\tau} + \frac{N(N-1)}{2}\frac{\Delta t_I}{\tau}. \tag{33}$$

For the timestep ratios we have used the reduction $r/r_{\text{c}}$ in computer time from using individual timesteps will lie between 0.28 (if $\Delta t_K \gg \Delta t_I$) and 0.46 ($\Delta t_K \ll \Delta t_I$); we actually obtained a reduction of 0.43.

We offer no simple prescription for choosing the timesteps $\tau_i$. The timesteps used in the sample integration were chosen by trial and error. Artificial stepsize resonances (Wisdom & Holman 1992) will be more numerous if there are individual timesteps, so comparison of two integrations with different timestep sets is always prudent.

We have not addressed errors arising through roundoff, which can be significant in long integrations. The usual methods for controlling roundoff

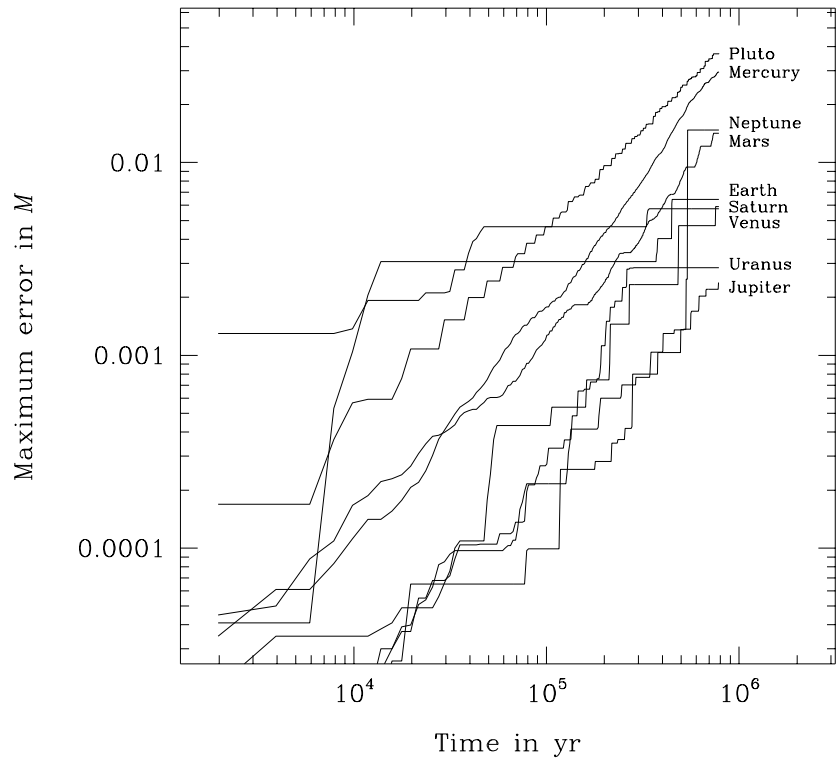

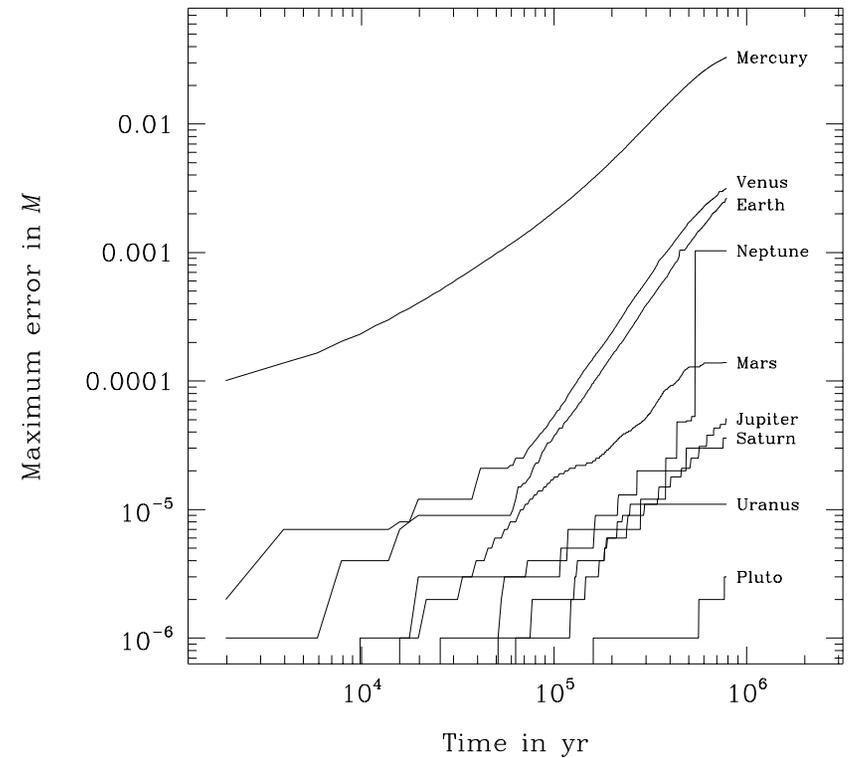

**1.** Plotted here is the maximum error in mean anomaly $M$ up to time $T$, against $T$, for leapfrog with individual timesteps. Mercury's timestep is $7\frac{1}{32}$ days, and the other timesteps are larger in the ratios $1:2:2:4:8:8:64:64:256$. The integration includes effects from general relativity and the Moon but does not integrate the lunar orbit. The faster-than-linear error growth for Mercury (noticeable for all the inner planets in Fig. 2) is presumably due to roundoff error.

error, based in part on carrying out selected additions in quadruple precision (Quinn & Tremaine 1990), are not effective in MVS integrators, where most of the roundoff arises during the repeated conversion between Cartesian coordinates and Kepler elements.

**2.** Like Fig. 1, but with all the planets having a timestep of $7\frac{1}{32}$ days. Note that the vertical scale has been shrunk to accommodate the much larger range of accuracies.

## 7. VARIANTS

Many variants are possible on the integrators we have discussed. A trivial and well-known example is to interchange the roles of the $K$ and $I$ clocks in the leapfrog cycle (10). For the two-timestep integrator (15), an alternative with similar properties is:

$\langle$ Advance $K_1$ by $\frac{1}{2}\tau$ $\rangle$

$\langle$ Advance $I$ by $\tau$ $\rangle$

$\langle$ Advance $K_1$ by $\tau$ and $K_2$ by $2\tau$ $\rangle$ \hfill (34)

$\langle$ Advance $I$ by $\tau$ $\rangle$

$\langle$ Advance $K_1$ by $\frac{1}{2}\tau$ $\rangle$

In contrast to simple leapfrog, there is no time-reversible algorithm of equal simplicity to (15) or (34) that begins by advancing the clock $I$.

The methods described here can be used to integrate other Hamiltonians of the form (11) that satisfy the Poisson bracket relations (12). Our methods can even be applied to systems that do not satisfy the relations (12b): simply make the assignments $H_{\text{int},1} \leftarrow H_{\text{int}}$, $H_{\text{int},i} \leftarrow 0$, $i = 2..N$.

As an example of an application in another physical context, we consider the integrator for use in molecular dynamics that is described by Skeel & Biesiadecki (1994). They wish to follow the motion of a particle under the actions of forces $F_1$ and $F_2$, where $F_1$ varies more rapidly than $F_2$ and the timestep for evaluating $F_2$ is $M$ times the timestep for $F_1$ (for simplicity we examine only the case $M = 2$). We can derive their algorithm from (15) by re-defining the clocks $K_i$ and $I$: when $K_i$ is running the particle's coordinates stay fixed while its momentum changes according to the force $F_i$, and while $I$ is running its position changes at fixed momentum. For a particle of unit mass and phase-space coordinates $(x, v)$ we then have

$\langle$ Increment $v$ by $\frac{1}{2}\tau F_1 + \tau F_2$ $\rangle$

$\langle$ Increment $x$ by $\tau v$ $\rangle$

$\langle$ Increment $v$ by $\tau F_1$ $\rangle$ \hfill (35)

$\langle$ Increment $x$ by $\tau v$ $\rangle$

$\langle$ Increment $v$ by $\frac{1}{2}\tau F_1 + \tau F_2$ $\rangle$

which is the integrator given by Skeel & Biesiadecki.

We thank Pierre-Vincent Koseleff and Ken Wilson for discussions, and Mark Bartelt for advice on code optimization. This research was supported by NSERC.